\documentclass[12pt]{article}
\usepackage{amsmath}
\usepackage{epsf}
\usepackage{epsfig}
\usepackage{amssymb}
\usepackage{graphicx}
\usepackage{latexsym}
\newcommand{\be}{\begin{equation}}
\newcommand{\ee}{\end{equation}}
\newcommand{\bear}{\begin{eqnarray}}
\newcommand{\eear}{\end{eqnarray}}

\newcommand{\GeV}{\mbox{GeV}}

\newsavebox{\LSIM}
\sbox{\LSIM}{\raisebox{-1ex}{$\ \stackrel{\textstyle<}{\sim}\ $}}

\newsavebox{\GSIM}
\sbox{\GSIM}{\raisebox{-1ex}{$\ \stackrel{\textstyle>}{\sim}\ $}}
\newcommand{\gsim}{\usebox{\GSIM}}
\newcommand{\scs}{\scriptsize}
\newcommand{\nn}{\nonumber}

\begin{document}
\begin{titlepage}
\begin{flushright}
CERN-PH-TH/2006-094\\
BI-TP 2006/18\\
hep-ph/0605242
\end{flushright}
$\mbox{ }$
\vspace{.1cm}
\begin{center}
\vspace{.5cm}
{\bf\Large Baryogenesis in the Two-Higgs Doublet Model}\\[.3cm]
\vspace{1cm}

Lars Fromme$^{a,}$\footnote{fromme@physik.uni-bielefeld.de},
Stephan J.~Huber$^{b,}$\footnote{stephan.huber@cern.ch}
and 
Michael Seniuch$^{a,}$\footnote{seniuch@physik.uni-bielefeld.de} \\ 
 
\vspace{1cm} {\em  
$^a$Fakult\"at f\"ur Physik, Universit\"at Bielefeld, D-33615 Bielefeld, Germany
}\\[.2cm] 
{\em $^b$Theory Division, CERN, CH-1211 Geneva 23, Switzerland}

\end{center}
\bigskip\noindent
\vspace{1.cm}
\begin{abstract}
We consider the generation of the baryon asymmetry in the two-Higgs doublet model.
Investigating the thermal potential in the presence of CP violation, as relevant for
baryogenesis, we find a strong first-order phase transition if the extra Higgs states
are heavier than about 300 GeV. The mass of the lightest Higgs can be as large 
as about 200 GeV.
We compute the bubble wall properties, including the profile of the relative complex phase
between the two Higgs vevs. The baryon asymmetry is generated by top transport,
which we treat in the WKB approximation. We find a baryon asymmetry consistent
with observations. The neutron electric dipole moment is predicted to be larger than
about $10^{-27}\;e\,{\rm cm}$ and can reach the current experimental bound.
Low values of $\tan\beta$ are favored.
\end{abstract}
\end{titlepage}
\section{Introduction}
The origin of the baryon asymmetry of the universe (BAU) is still an open
question in cosmology and particle physics. New measurements of the cosmic
microwave background, combined with large-scale structure data, 
yield a baryon to entropy ratio of \cite{CMB}
\be
\eta_B\equiv\frac{n_B}{s}=(8.7\pm 0.3)\times 10^{-11}.
\ee
Three necessary conditions, stated by Sakharov \cite{Sakharov67}, have to be
fulfilled for a dynamical generation of the baryon asymmetry: baryon number
violation, C and CP violation, and departure from thermal equilibrium. In
principle the standard model (SM) contains all these requirements, and the
electroweak phase transition (EWPT) provides a natural mechanism for
baryogenesis \cite{Kuzmin85}. The baryon asymmetry is generated during the
phase transition by electroweak sphaleron processes. To avoid subsequent 
baryon number washout, the sphaleron rate has to be suppressed after the 
phase transition.
Hence the transition must be strongly first order, i.e.\ the expectation value of the
Higgs field must be larger than about the critical temperature. 
In the SM there is no first-order phase transition for Higgs masses larger than
about 80 GeV \cite{KLRS96}, 
far below the current experimental bound of $m_H>114$~GeV \cite{m_H}.
The SM therefore fails to explain the baryon asymmetry. Moreover, the CP violation in the 
CKM matrix is too small to produce a sufficiently large baryon number \cite{FS93}.

Over the years there have been many proposals to realize electroweak
baryogenesis in extended models (see, for instance, ref.~\cite{review} for a review). 
In supersymmetric theories, for example, 
a strong first-order PT can occur if the superpartner of the top quark is
lighter than about 175 GeV \cite{MSSM}, and the baryon asymmetry can be
generated by chargino transport \cite{CJK00,MSSM1}. Alternatively, the phase transition
can be strengthened by the presence of SM singlets in the Higgs sector
\cite{NMSSM}. 
A more general effective field theory approach can also be followed; there the
Higgs sector is augmented by dimension-six operators to induce a first order
phase transition and to provide additional CP violation
\cite{eff,BFHS05,FH06}.

In this paper we revisit electroweak baryogenesis in the two-Higgs doublet model 
(2HDM) \cite{BKS90,CKN91,TZ91,FKT93,DFJM94,CKV95,CL96}. In addition
to the SM Higgs, the 2HDM contains two extra neutral and charged Higgs
particles. If these extra states couple sufficiently strongly, their thermal loop
corrections can induce a strong first-order phase transition \cite{BKS90,TZ91,DFJM94,CL96}.
In addition, a complex mass term, mixing the two Higgs doublets, provides a new
source of CP violation, which can fuel baryogenesis. 

We examine the EWPT in the 2HDM, including explicit CP violation, using the 
finite temperature effective potential at one-loop order. 
In agreement with ref.~\cite{CL96},
we find a strong phase transition for light Higgs masses of up to at least 200 GeV.
The extra Higgses have to be heavier than about 300 GeV, depending somewhat
on the model parameters. Turning on the CP-violating phase makes the phase
transition slightly weaker. We determine the profile of the bubble wall, which 
separates the broken and symmetric phase. Except for the case of very strong
phase transitions, we typically find thick bubble walls. The bubble wall is
characterized by a varying complex phase between the two Higgs vevs. 
CP-violating interactions of the particles in the hot plasma, in particular the
top quark, with the phase boundary then lead to different semiclassical forces
acting on particles and antiparticles. Since we are dealing with thick bubble
walls, we can apply the standard WKB formalism to compute the CP-violating
source terms that enter the transport equations of electroweak baryogenesis
\cite{CJK00,JPT}.
Here we use the formalism recently laid out in ref.~\cite{FH06}, which makes sure
that the correct dispersion relations of the Schwinger--Keldysh formalism \cite{PSW1}
are reproduced. Also a finite $W$-scattering rate is included in the transport
equations, which previously was set to equilibrium.

We find that a wide parameter range allows for the generation the observed baryon
asymmetry. Since the model contains only a single CP-violating phase, we can
predict the electric dipole moments of the neutron and electron. They are
typically found to be below the current experimental bounds, but should be
detectable in next-generation experiments.

\section{The effective Higgs potential}
In its most general form, the 2HDM suffers from flavor changing
neutral currents at tree-level. To avoid this, a discrete symmetry,
$H_1\rightarrow-H_1$, $d_i^c\rightarrow \mp d_i^c$ (the other fields do not transform),
is usually invoked \cite{GW77}, making sure that at most one Higgs doublet couples
to the up- and down-type quarks, respectively.
In the ``$-$'' case (``type II''), the down-type quarks
couple only to $H_1$, while the up-type ones couple to $H_2$. 
In the other case (``type I''), $H_1$ does not couple to the fermions at all.
In the following only the coupling of the top quark will be relevant, so that
we do not need to actually distinguish between types I and II. In section 4, where we
will discuss electric dipole moments, we will focus on the type II case.  

The most general potential is \cite{S89}
\bear
\label{V0}
V_0(H_1,H_2)&=&
-\mu_1^2 H_1^\dagger H_1-\mu_2^2 H_2^\dagger H_2
-\mu_3^2(e^{i\phi} H_1^\dagger H_2+{\rm h.c.})\nonumber\\
&&+\frac{\lambda_1}{2}(H_1^\dagger H_1)^2+\frac{\lambda_2}{2}(H_2^\dagger H_2)^2
+\lambda_3(H_1^\dagger H_1)(H_2^\dagger H_2)\nonumber\\
&&+\lambda_4|H_1^\dagger H_2|^2
+\frac{\lambda_5}{2}\left((H_1^\dagger H_2)^2+{\rm h.c.}\right)
\eear
and the Yukawa interactions read
\be \label{Yuk}
{\cal L}_y=yH_2Q_3t^c+\mbox{h.c.}+\dots
\ee
Without loss of generality the couplings $\lambda_i$ and the mass parameters
$\mu_i$ can taken to be real. The mass term $\mu_3^2e^{i\phi}$
breaks the aforementioned
$Z_2$ symmetry softly, without reintroducing tree-level flavor violation \cite{BR85}.
It can be complex, in which case the Higgs potential breaks CP. 
In total, the Higgs potential contains 9 parameters, which are
3 squared masses, 5 couplings, and 1 phase. One parameter is fixed
by the $Z$-boson mass, leaving an 8-dimensional parameter space. 
The potential has to be bounded from below, which at tree-level
translates into the constraints \cite{DFJM94}
\bear
\lambda_1>0,~\lambda_2>0, ~\sqrt{\lambda_1\lambda_2}+\lambda_3>0
, ~\sqrt{\lambda_1\lambda_2}+\lambda_3+\lambda_4\pm\lambda_5>0.
\eear

Let us first consider the case $\phi=0$, i.e. the Higgs potential without CP violation. 
It has been shown that in this case no charge breaking minima exist, provided the
charged Higgs mass squared is positive \cite{FSB04}\footnote{A study of
symmetry breaking in a general 2HDM has recently been presented in ref.~\cite{MMNN06}.}. 
Later on we will assume that
this result generalizes to the one-loop level, including a small CP-violating phase.
We can therefore restrict ourselves to the neutral fields, which we parameterize as
${\rm Re}H_1^0=h_1$ and ${\rm Re}H_2^0=h_2$. The potential then reads
\be
V_0(h_1,h_2)=-\mu_1^2h_1^2-\mu_2^2h_2^2-2\mu_3^2h_1h_2
+\frac{\lambda_1}{2}h_1^4+\frac{\lambda_2}{2}h_2^4
+(\lambda_3+\lambda_4+\lambda_5)h_1^2h_2^2.
\ee
In the following we focus on the somewhat simpler case
\be \label{eq}
\mu_1^2=\mu_2^2,~\lambda_1=\lambda_2.
\ee
Moreover, this choice is favorable to generate large Higgs
expectation values in the broken phase \cite{BKS90}.
The Yukawa interaction (\ref{Yuk}) does not preserve these relations 
at the loop-level. At tree-level, eq.~(\ref{eq}) implies the symmetry
\be \label{sym}
H_1\leftrightarrow H_2^{\dagger},
\ee
so that the minimum is at $\tan\beta\equiv\langle h_2\rangle/\langle h_1\rangle=1$.
With $\langle h_1\rangle=\langle h_2\rangle=h=123$~GeV the extremal condition is
then given by
\be
-\mu_1^2-\mu_3^2+(\lambda_1+\lambda_3+\lambda_4+\lambda_5)h^2=0.
\ee
The mass matrix is block-diagonal and we obtain, besides 3 massless Goldstone
bosons, 5 physical Higgs bosons. They consist of a pair of charged Higgses 
$H^\pm$, 2 neutral scalars $h^0$ and $H^0$, and a pseudoscalar $A^0$, 
with the corresponding squared masses as follows: 
\bear
m^2_{H^\pm}&=&2\mu_3^2-2(\lambda_4+\lambda_5)h^2,\\
m^2_{A^0}&=&2\mu_3^2-4\lambda_5h^2,\\
m^2_{H^0}&=&2\mu_3^2-2(-\lambda_1+\lambda_3+\lambda_4+\lambda_5)h^2,\\
m^2_{h^0}&=&2(\lambda_1+\lambda_3+\lambda_4+\lambda_5)h^2.
\eear 
These relations can be used to define the model in terms of $\mu_3^2$
and the 4 Higgs masses.

In the case of non-vanishing $\phi$, CP is broken. We now parametrize the 
neutral Higgs fields as
\be\label{neutral_field_para}
H_1^0=h_1e^{-i\theta_1},~H_2^0=h_2e^{i\theta_2}.
\ee
Note that the potential only depends on the combination $\theta=\theta_1+\theta_2$. 
In the minimum we can always choose the gauge such that $\theta_1=\theta_2=\theta/2$. 
Still assuming the relations~(\ref{eq}), the potential of the neutral fields reads
\bear
V_0(h_1,h_2,\theta)&=&-\mu_1^2(h_1^2+h_2^2)-2\mu_3^2h_1h_2\cos(\theta+\phi)
+\frac{\lambda_1}{2}(h_1^4+h_2^4)
\nonumber\\
&&+(\lambda_3+\lambda_4+\lambda_5\cos(2\theta))h_1^2h_2^2.
\eear
Using the notation
$\langle\theta\rangle=\vartheta$, we obtain two extremal conditions,
\bear
-\mu_1^2-\mu_3^2\cos(\vartheta+\phi)
+(\lambda_1+\lambda_3+\lambda_4+\lambda_5\cos(2\vartheta))h^2&=&0
\nonumber\\
\mu_3^2\sin(\vartheta+\phi)-\lambda_5\sin(2\vartheta)h^2&=&0.
\eear
The squared Higgs boson masses take the form
\bear \label{Hmass}
m^2_{H^\pm}&=&-2\mu_1^2+2(\lambda_1+\lambda_3)h^2,\nonumber\\
m^2_{H_3}&=&-\mu_1^2+2(\lambda_1+\lambda_3+\lambda_4)h^2+\sqrt{\mu_1^4+4\lambda_5\cos(2\vartheta)\mu_1^2h^2+4\lambda_5^2h^4},\nonumber\\
m^2_{H_2}&=&-2\mu_1^2+4\lambda_1h^2,\nonumber\\
m^2_{H_1}&=&-\mu_1^2+2(\lambda_1+\lambda_3+\lambda_4)h^2-\sqrt{\mu_1^4+4\lambda_5\cos(2\vartheta)\mu_1^2h^2+4\lambda_5^2h^4}.
\eear
Note that the neutral Higgs states are now mixtures, with scalar and pseudoscalar contents.
Again, these relations can be inverted to parameterize the model in terms of the Higgs 
masses, $\mu_3^2$ and $\phi$.

At zero temperature the one-loop contribution to the effective potential is given by
\be
V_1(H_1,H_2)=\sum_i\pm\frac{n_i}{64\pi^2}m_i^4\ln\frac{m_i^2}{Q^2},
\ee
where $m_i^2=m_i^2(H_1,H_2)$ are field dependent mass eigenvalues, $n_i$ is the
corresponding number of degrees of freedom, and "$+ (-)$"
applies to bosonic (fermionic) contributions, respectively. We choose
$Q=246/\sqrt{2}$ GeV for the renormalization scale. We take only the heaviest
bosons, i.e. $m_i = m_{H^\pm}, m_{H_2}, m_{H_3}$ ($n_{H^\pm}=n_{H_2}=n_{H_3}=1$),
and the
fermion with the largest Yukawa coupling, i.e. $m_i = m_t$ ($n_t=12$), into
account. For the top quark mass we have $m_t^2=y_t^2H^\dagger_2H_2$. All
other particles can be safely neglected, owing to their small
contributions to the effective one-loop potential.

We add counter-terms to the potential, such that the tree-level minimum and 
Higgs masses are preserved at the one-loop level. This can be achieved by 
\bear
\label{VCT}
V_{\rm CT}(H_1,H_2)&=&
-\delta\mu_1^2 (H_1^\dagger H_1 + H_2^\dagger H_2)
-\delta\mu_3^2(e^{i\phi} H_1^\dagger H_2+{\rm h.c.})\nonumber\\
&&+\frac{\delta\lambda_1}{2}(H_1^\dagger H_1)^2+\frac{\delta\lambda_2}{2}(H_2^\dagger H_2)^2
+\delta\lambda_3(H_1^\dagger H_1)(H_2^\dagger H_2)
\nonumber\\
&&+\delta\lambda_4|H_1^\dagger H_2|^2
+\frac{\delta\lambda_5}{2}\left((H_1^\dagger H_2)^2+{\rm h.c.}\right).
\eear
As already mentioned, the symmetry
(\ref{sym}) no longer holds at the one-loop level, which we take care of
by using  $\delta\lambda_1\neq\delta\lambda_2$. 
Three renormalization conditions are evidently given by
\be
\left.\frac{\partial(V_1+V_{\rm CT})}{\partial
    h_1}\right|_{h_1=h_2=h\atop\theta=\vartheta\hfill}=\left.\frac{\partial(V_1+V_{\rm CT})}{\partial
    h_2}\right|_{h_1=h_2=h\atop\theta=\vartheta\hfill}=\left.\frac{\partial(V_1+V_{\rm CT})}{\partial
    \theta}\right|_{h_1=h_2=h\atop\theta=\vartheta\hfill}=0,
\ee
meaning that the minimum of the potential $V=V_0+V_1+V_{\rm CT}$ does not change with 
respect to the tree-level case. Preserving the values of the Higgs masses,
which we compute from the second derivatives of $V$,
provides another four conditions.  
So the coefficients of $V_{\rm CT}$ are fixed. Since the conditions related to Higgs
masses include non-linearities, the resulting equations for the counter-terms
have to be solved numerically.

The 2HDM is subject to a number of experimental constraints. In the considered
parameter range, the lightest Higgs boson is SM-like, and therefore has
to obey the LEP bound of $m_H>114$~GeV \cite{m_H}.
The 2HDM does not respect the custodial symmetry of the SM. So there is
the danger of large corrections to the electroweak precision observables. 
These corrections can be approximately described in terms of contributions
to the self-energies, the so called ``oblique'' corrections. The relevant expressions
for the 2HDM with CP-violation can be found in ref.~\cite{FMK92}. To be consistent
with observations, the mass splittings between the extra Higgs states should not
be much larger than the W-mass. Later on we will set these masses equal to reduce
the number of parameters. Oblique corrections then are automatically small.
Another important constraint comes from $b\rightarrow s\gamma$, which in the
type II model requires $m_{H^\pm}\gsim200$~GeV \cite{N04}\footnote{One can
also consider constraints from $B_d^0-\bar B_d^0$ mixing \cite{Bd06}.}. Constraints 
from the muon anomalous magnetic moment \cite{CK03} and from tau decays \cite{KT04}
are not relevant for the low values of $\tan\beta$, which we consider.

At finite temperature the one-loop contribution to the effective potential is given by
\be\label{VeffT}
V_1^T=T^4 \sum_B n_B\; f_B\left(\frac{m_B}{T}\right) + T^4 \sum_F n_F\; f_F\left
(\frac{m_F}{T}\right),
\ee
where $n_{B(F)}$ counts the positive degrees of freedom for bosons (fermions). 
In the high temperature limit, $m/T\ll 1$, one obtains \cite{DJ74}
\bear\label{fBHT}
f_B^{\rm HT}\left(\frac{m}{T}\right)&\approx&\frac{-\pi^2}{90}+\frac{m^2}{24T^2}
 -\frac{m^3}{12\pi T^3}
 -\frac{m^4}{64\pi^2T^4}\ln\left(\frac{m^2}{c_BT^2}\right)\\
f_F^{\rm HT}\left(\frac{m}{T}\right)&\approx&\frac{-7\pi^2}{720}+\frac{m^2}{48T^2}
 +\frac{m^4}{64\pi^2T^4}\ln\left(\frac{m^2}{c_FT^2}\right)\label{fFHT},
\eear
with $c_F=\pi^2\exp(3/2-2\gamma_e)\approx 13.94$ and $c_B=16c_F$, and in the
low temperature limit, when $m/T$ is large,
\be\label{fLT}
f^{\rm LT}\left(\frac{m}{T}\right)
\approx-\left(\frac{m}{2\pi T}\right)^{3/2}\exp\left(-\frac{m}{T}\right)\left(1+\frac{15m}{8T}\right).
\ee
In this low temperature limit the contributions from bosons and fermions have the same
asymptotic behavior.

We use these approximations because they are much more convenient to handle than the
full integral expressions of ref.~\cite{DJ74}. It turns out, however, that these 
limiting cases are not sufficient since some states cross from the high temperature
to the low temperature regime.  
For an expression to be valid in the whole temperature range, we therefore use
a smooth interpolation between the low- and high-$T$ limits. For bosons we
use eq.~(\ref{fBHT}) for $m/T<1.8$ and eq.~(\ref{fLT}) for $m/T>4.5$, and for 
fermions eq.~(\ref{fFHT}) for $m/T<1.1$ and eq.~(\ref{fLT}) for $m/T>3.4$. 
The interpolations are made in such a way that
the functions as well as their derivatives match at the connecting points. The
deviation between our approximation and the exact
solution is less than 4\%.
Finally, the effective potential is given by
\be
V_{\mbox{\scs eff}}=V_0+V_1+V_{\rm CT}+V_1^T.
\ee
In $V_1^T$ we also take into account the contributions of $W$-bosons ($n_W=6$)
and $Z$-bosons ($n_Z=3$). In perturbation theory the strength of a strong phase
transition would be underestimated by resummation of the gauge boson contributions
\cite{LR01}. Therefore we do not resum these corrections to
compensate for non-perturbative effects.
In ref.~\cite{CL96} the thermal contributions 
of the heavy Higgs bosons have been resummed,
using the high temperature approximation for the thermal Higgs masses.
We find that in the broken minimum the high temperature approximation is often not
justified for the heavy Higgs states. Ignoring this fact, and nevertheless 
resumming the Higgs
contributions by using thermal masses instead of the bare masses in eq.~(\ref{VeffT}),
we find a phase transition less than 15\% weaker than in unresummed case. 
The effect is marginal for large heavy Higgs masses and becomes stronger for
smaller ones.
Including the light Higgs and the Goldstone bosons has an even smaller effect.
In the results we present below, the Higgs contributions are not resummed.

\section{The phase transition}\label{PT}
The dynamics of the EWPT is governed by $V_{\rm eff}(h_1,h_2,\theta,T)$. 
The critical temperature, $T_c$, of a 
first-order phase transition is defined by the condition that the effective potential
has two degenerate
minima, the symmetric minimum at $\langle h_1\rangle_T=\langle h_2\rangle_T=0$
and the broken minimum at $\langle h_1\rangle_T=v_1>0$
and $\langle h_2\rangle_T=v_2>0$, which are separated by an energy barrier.
The total Higgs expectation value we define by $v_c=\sqrt{2}\sqrt{v_1^2+v_2^2}$,
where the factor $\sqrt{2}$ is due to our normalization of the Higgs fields.
Somewhat below $T_c$, at the nucleation temperature $T_n$, bubbles of the
broken phase start to nucleate and expand. Baryogenesis takes place outside the bubbles
in the symmetric phase while, inside the bubbles, the sphaleron rate that
provides $(B+L)$-violating processes has to be switched off. Otherwise the
baryon asymmetry will be washed out after the phase transition. In order to
preserve the created baryon asymmetry, the washout criterion \cite{washout}
\be
\xi=\frac{v_c}{T_c}\gtrsim 1
\ee
must hold, i.e.~the phase transition has to be sufficiently strong.  

In the following we analyze the parameter space with respect to 
the strength $\xi$ of the phase
transition. We focus on the case of degenerate heavy Higgs masses, which
reduces the dimension of the parameter space. As input
parameters we take $\mu_3^2, \phi, m_h=m_{H_1},$ and
$m_H=m_{H_2}=m_{H_3}=m_{H^\pm}$. One finds that for larger values of
$\phi$ ($\phi=0.4$ for example) the first-order phase transition
can change into a two-stage transition if the heavy Higgs mass is sufficiently small.
The potential then shows an additional local minimum. The phase transition proceeds
by a second-order phase transition from the symmetric phase to this extra minimum, 
followed by a first-order phase transition to the low temperature broken phase\footnote{The 
second-transition is in general too weak to avoid baryon number washout. Also, the non-zero
Higgs vev outside the bubbles would lead to a suppression of the sphaleron
rate and therefore also of the baryon asymmetry. So the two-stage transition does not 
allow us to generate the baryon asymmetry.}.
We exclude these values from the parameter space and only define the strength
$\xi$ in the case of a pure first-order PT.

Another important property that enters the transport equations discussed in
chapter 5 is the wall profile of the expanding
bubbles. If the nucleating bubbles have reached a sizable extent and expand with
constant velocity, we can boost into the rest frame of the bubble wall and
assume a planar wall.
In principle one has to numerically solve the field equations of the Higgs fields,
using an algorithm such as the one recently proposed in ref.~\cite{KH06}.
To achieve a sufficiently strong phase transition we are led to $m_H^2\gg m_h^2$.
The effective potential is then characterized by a valley, corresponding to
the single light Higgs field. During the phase transition the fields will follow
this valley very closely, in order not to feel the heavy Higgs masses. So we
can approximate the phase transition by single field dynamics. Numerically
we determine the valley by minimizing the thermal potential at $T_c$ with
respect to $h_2$ and $\theta$ at fixed values of $h_1$ between the 
symmetric and broken phase.
For a simple $\varphi^4$ model, with one real scalar field $\varphi$ and a broken
minimum at $v_c$, the wall profile is exactly described by a kink solution,
\be
\varphi(z)=\frac{v_c}{2}\left(1-\mbox{tanh}\frac{z}{L_{\rm w}}\right),
\ee
with a wall thickness $L_{\rm w}=\sqrt{v_c^2/(8V_b)}$, where $V_b$ is the 
height of the potential barrier
and $z$ is the coordinate orthogonal to the wall.
\begin{figure}[p]
\begin{center} 
   \epsfig{file=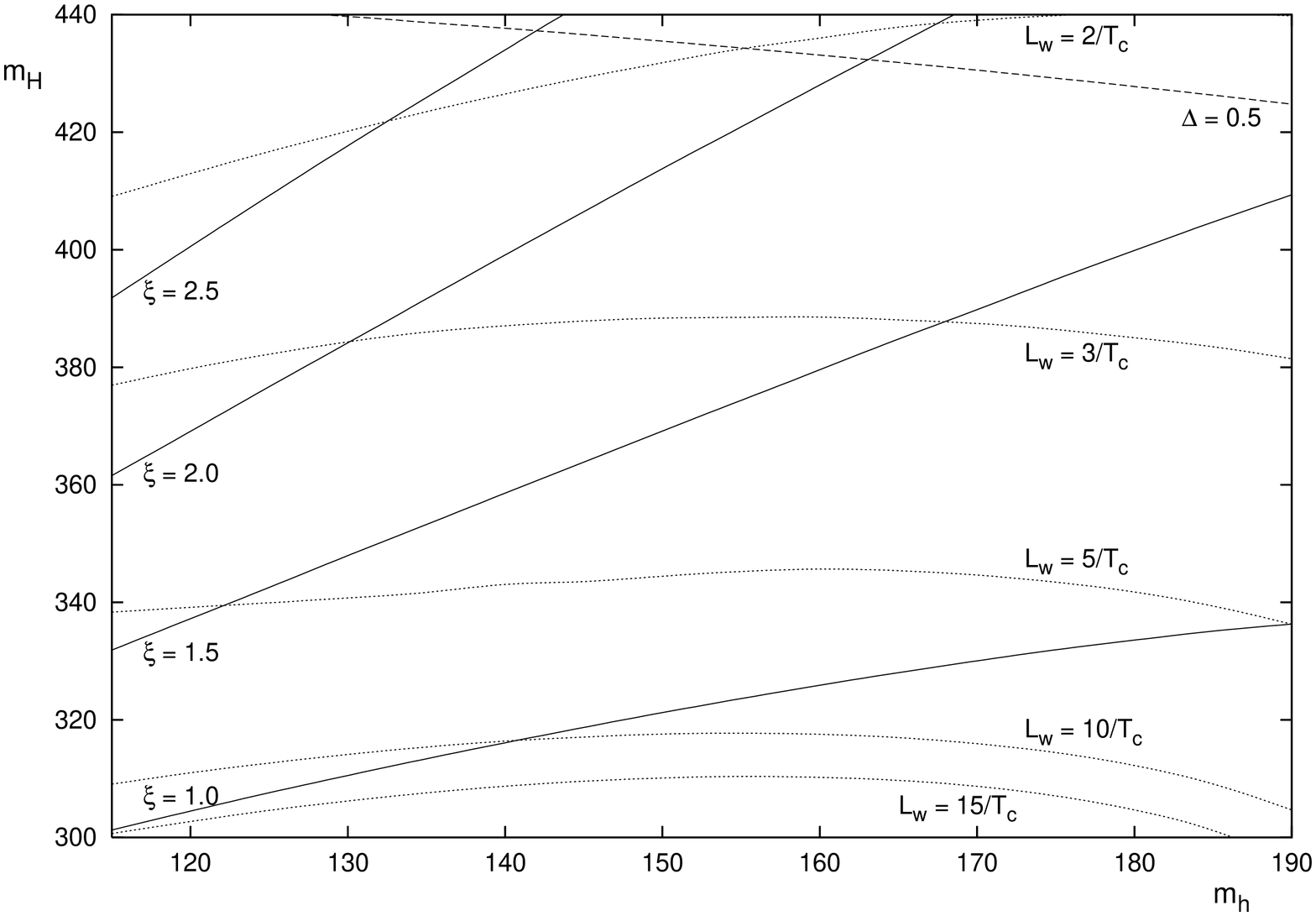,width=85mm,angle=270}
\end{center}
\caption{Lines of constant $\xi$ and $L_{\rm w}$ in the $m_h$-$m_{
    H}$-plane for $\mu_3^2=10000~\GeV^2$ and $\phi=0$. In addition, the line of
    the relative size of the one-loop corrections
    $\Delta=\max|\delta\lambda_i/\lambda_i|=0.5$ is 
    shown. The Higgs masses are given in units of GeV.}
\label{fig_10000_0.0001}
\begin{center} 
   \epsfig{file=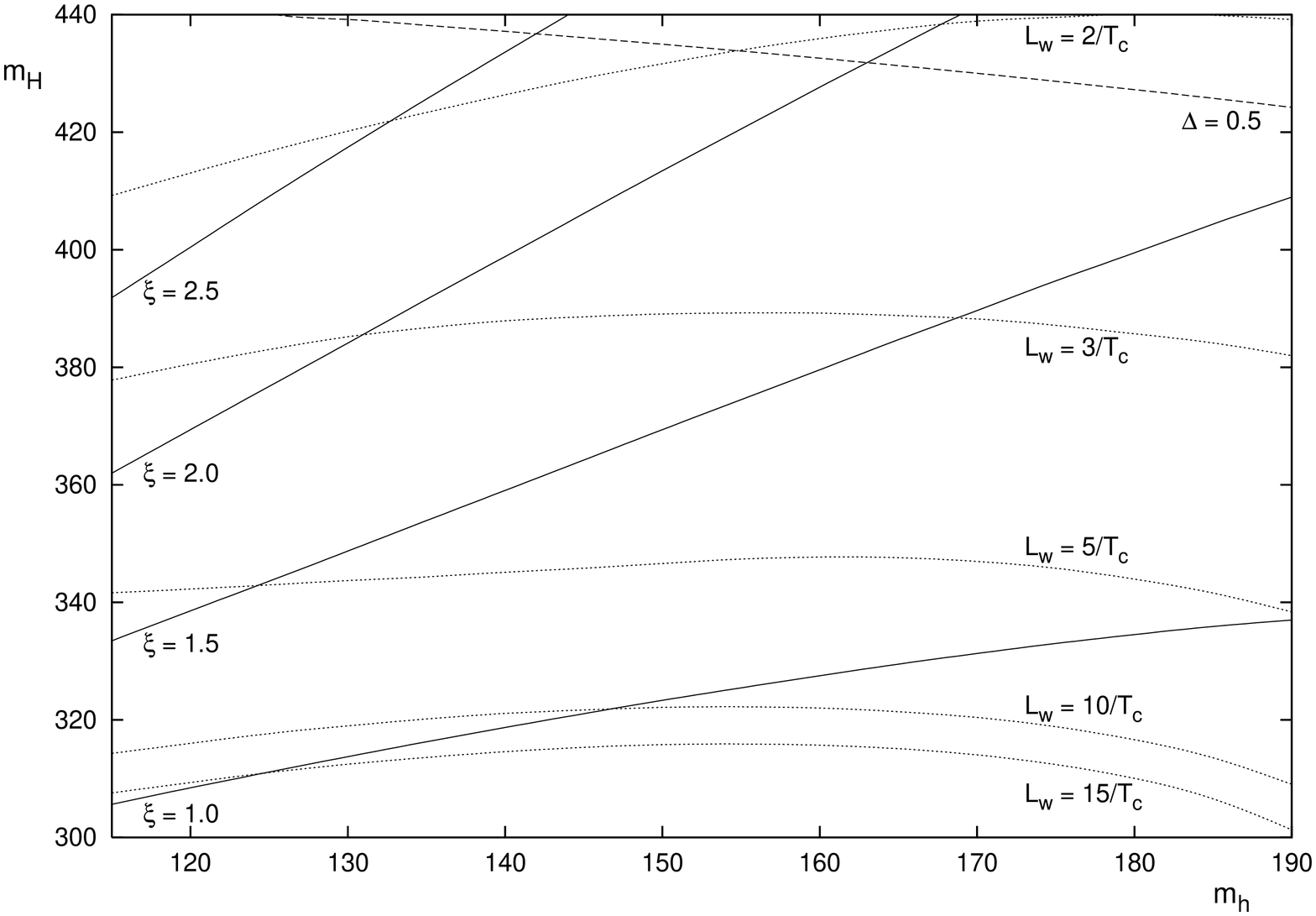,width=85mm,angle=270}
\end{center}
\caption{The same plot as in fig.~\ref{fig_10000_0.0001}, but for the set
  $\mu_3^2=10000~\GeV^2$ and $\phi=0.2$.} 
\label{fig_10000_0.2}
\end{figure}
We use this approximation for $L_{\rm w}$, determining $V_b$ 
as the maximal height of the potential barrier along the valley connecting the
two minima.
 
Let us now briefly discuss the behavior of $L_{\rm w}$ and $\xi$ with the input
parameters. 
We require $m_h\ge 115~\GeV$ to be consistent with the LEP bound on the 
Higgs mass.
For increasing $m_H$ and keeping the other parameters fixed, the wall 
thickness decreases, while the PT becomes stronger. This somewhat counter-intuitive
result is due to the fact that the larger Higgs masses (\ref{Hmass}) come from larger
quartic couplings. So this limit actually does not lead to the decoupling of the heavy
states. At some point perturbation theory will finally break down. Later on, when 
computing the baryon asymmetry, we will face another constraint. The gradient
expansions is justified only for thick bubble walls, so we will require
$L_{\rm w}T_c>2$.
In practice this leads to an upper bound on  $m_H$ similar to
the perturbativity constraint.

\begin{figure}[t]
\begin{center} 
   \epsfig{file=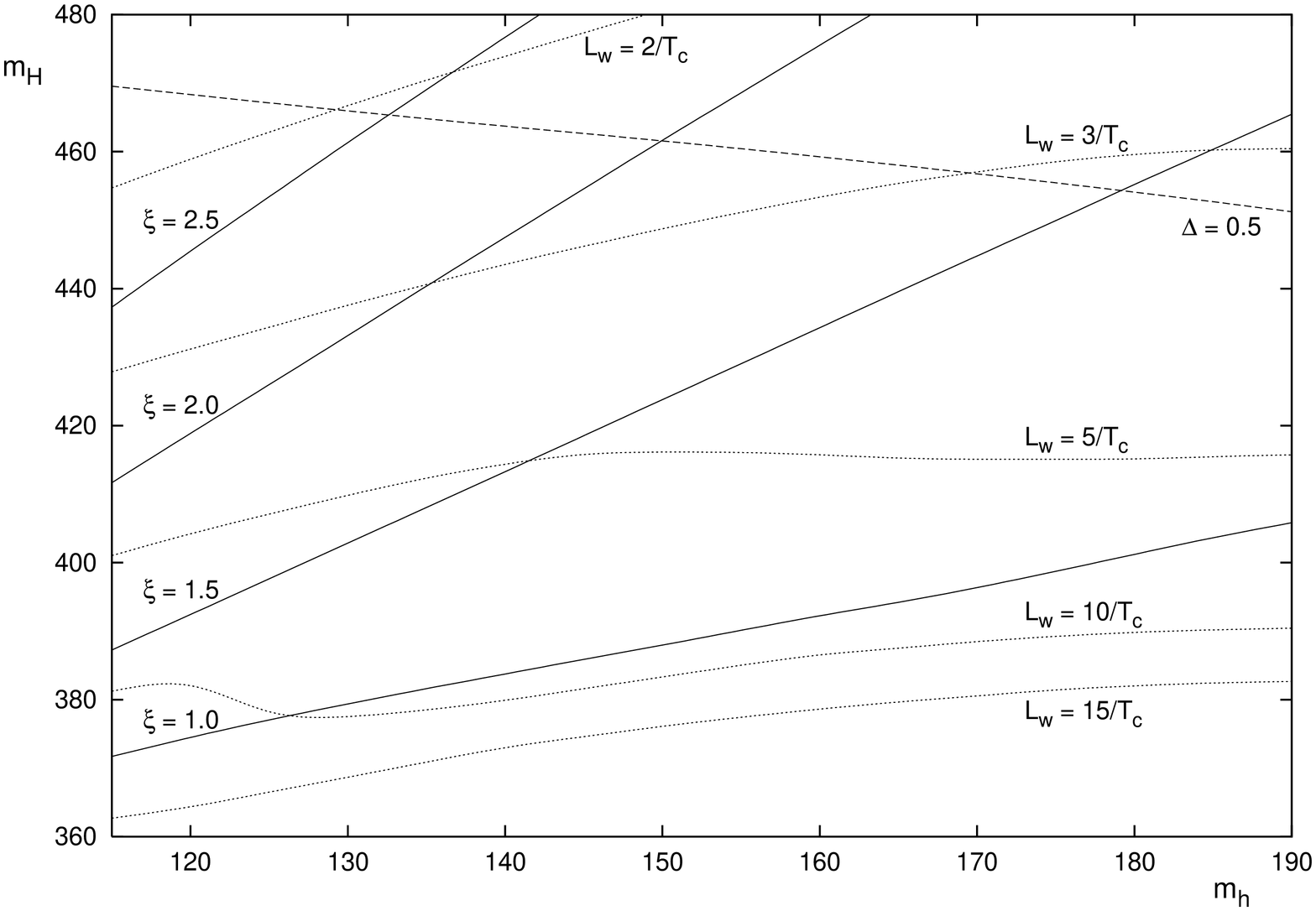,width=85mm,angle=270}
\end{center}
\caption{The same plot as in figs.~\ref{fig_10000_0.0001} and
  \ref{fig_10000_0.2}, but for $\mu_3^2=20000~\GeV^2$ and $\phi=0.2$}
\label{fig_20000_0.2}
\end{figure}

In figures \ref{fig_10000_0.0001}--\ref{fig_20000_0.2} constant
lines of $\xi$ and $L_{\rm w}$ 
are shown in the dependence of $m_h$ and $m_H$ for different
values of $\mu_3^2$ and $\phi$. The influence of the CP-violating phase $\phi$
on $\xi$ and $L_{\rm w}$
is rather small, as can be seen from figs.~\ref{fig_10000_0.0001} and
\ref{fig_10000_0.2}. For small 
values of $\xi$, or large values of $L_{\rm w}$, the lines are  marginally shifted 
upwards. This behavior continues for increasing $\phi$. If we choose $\phi=0.4$
the effect is negligible above $\xi\approx 1.5$, but below $\xi\approx 1.3$ the
PT changes into a two-stage one. 
The effect of increasing $\mu_3^2$ is a shift to higher values for $m_H$.
The comparison of $\mu_3^2=10000~\GeV^2$ and $20000~\GeV^2$, 
e.g.~figs.~\ref{fig_10000_0.2} and \ref{fig_20000_0.2}, shows that the 
range of $m_H$ is moved to higher values, while the extent shrinks
by around $20~\GeV$. Using a larger value of $\mu_3^2$ means that the
same quartic couplings lead to heavier Higgses. The strength of the phase
transition is more governed by the size of the quartic couplings than by the
actual value of  $m_H$. Notice that here the bound on the charged Higgs
mass from $b\rightarrow s\gamma$, which we discussed above, is automatically
satisfied in the case of a strong phase transition.

In figs.~\ref{fig_10000_0.0001} and \ref{fig_10000_0.2} the one-loop corrections
$\Delta=\max|\delta\lambda_i/\lambda_i|$ to the quartic couplings, i.e.~the size
of the counter-terms with respect to 
the tree-level terms, range from 15\% for $m_H=300$~GeV to
50\% for $m_H=440$~GeV. This means that in the case of a very strong 
phase transition, perturbation theory starts to break down and sizable corrections
to our results have to be expected. For $\xi\sim1$ higher-order corrections are
well under control. These results agree with the findings of ref.~\cite{CL96}.
In conclusion, we find that a wide range of parameters
fulfills the requirements of electroweak baryogenesis. 

\begin{figure}[t]
\begin{center} 
 \epsfig{file=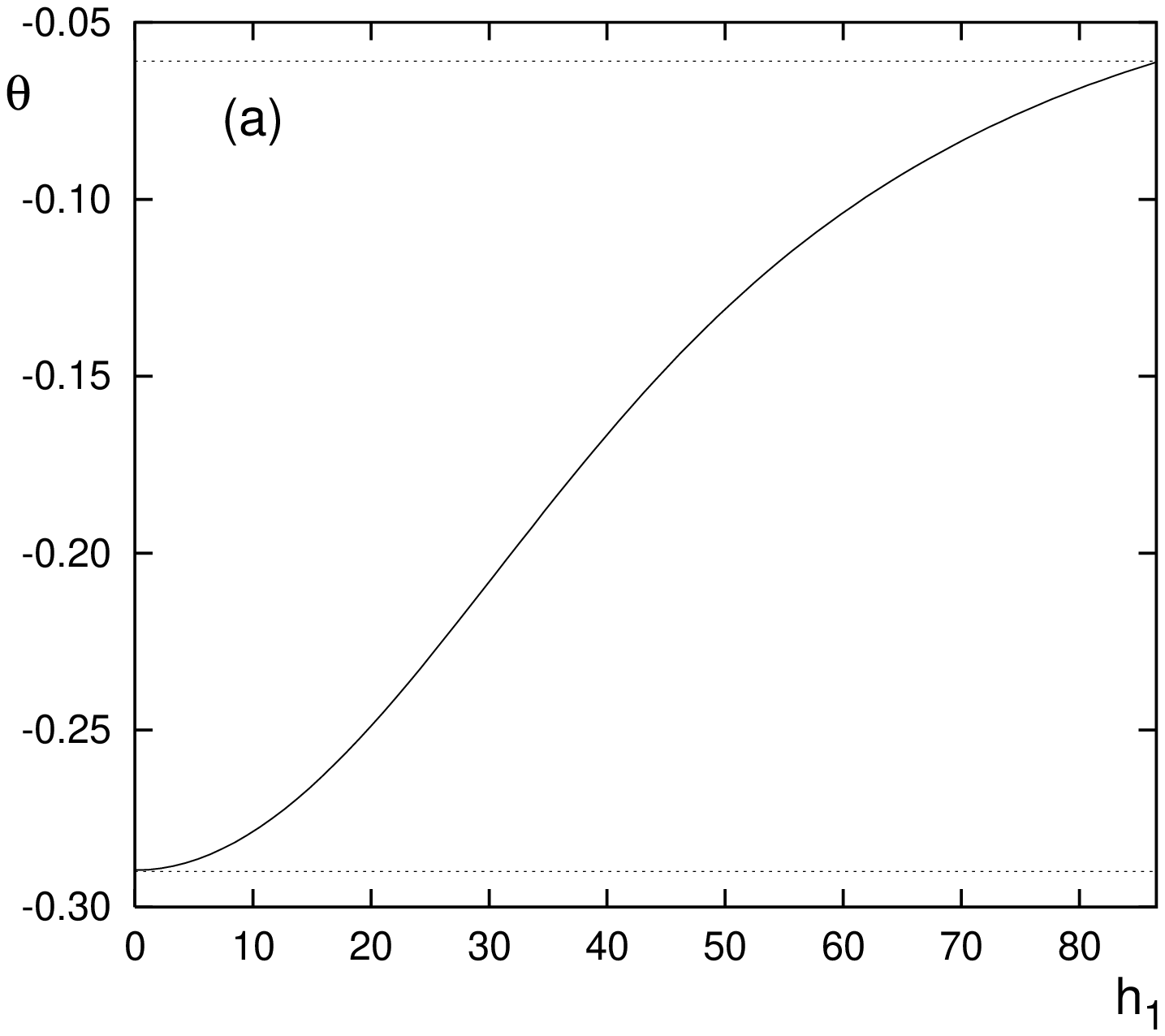,width=60mm,angle=270}
 \hfill\epsfig{file=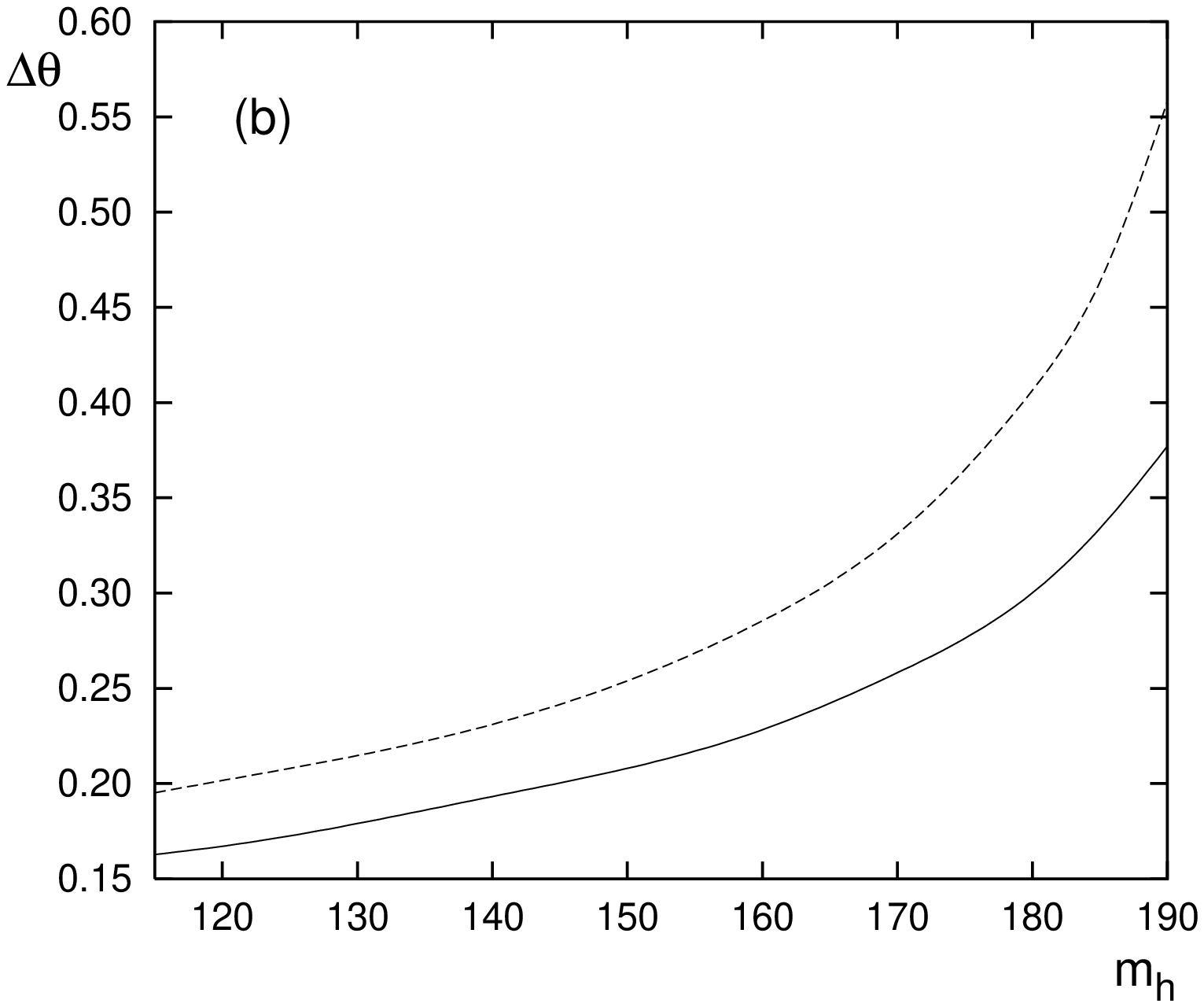,width=60mm,angle=270}
\end{center}
\caption{The phase $\theta$ and the difference $\Delta\theta$ for the set
 $\mu_3^2=10000~\GeV^2$, $\phi=0.2$.\newline
 (a) The change of $\theta$ during the PT, as a function of $h_1$ (given in GeV),
 at fixed $m_h=150\;\GeV$, $m_H=350\;\GeV$.\newline
 (b) $\Delta\theta$ versus $m_h$ (given in GeV) for $m_H=330\;\GeV$ (solid) and
 $400\;\GeV$ (dashed).}
\label{fig_theta_h2}
\end{figure}

Up to now we have only discussed observables involving the fields $h_1$ and $h_2$.
However, also the CP-violating phase $\theta$, which varies along the bubble wall from
$\theta_{\rm sym}$ to $\theta_{\rm brk}$, is essential for baryogenesis. According
to the above discussion, we compute the $\theta$-profile approximately
by minimizing the thermal potential at $T_c$ with respect to $h_2$ and $\theta$ 
at fixed values of $h_1$ between the symmetric and broken phase. 
As a representative example we show in 
fig.~\ref{fig_theta_h2}a the $\theta$-profile parametrized by $h_1$ for the
set $\mu_3^2=10000~\GeV^2$ and $\phi=0.2$. There, $\theta$ changes from
$\theta_{\rm sym}=-0.29$ to $\theta_{\rm brk}=-0.06$, which is indicated by the
dotted lines. In a simplified manner, we describe the 
$\theta$-profile by a kink ansatz, i.e.
\be
\theta(z)=\theta_{\rm brk}-\frac{\Delta\theta}{2}
\left(1+\tanh\frac{z}{L_{\rm w}}\right),
\ee
using the derived wall thickness $L_{\rm w}$ and
$\Delta\theta=\theta_{\rm brk}-\theta_{\rm sym}$. The CP violation in the
Higgs sector gives rise to complex fermion masses, which change while the
particles pass through the bubble wall. We only take into account the top quark as
the heaviest fermion. With our parametrization (\ref{neutral_field_para}) of
the neutral Higgs components, one finds for the complex top mass
\be \label{M}
{\cal M}_t(z)=y_t h_2(z)\,e^{i\theta(z)/2}=y_t\frac{h(z)}{\sqrt{2}}
\sin\beta_T\,e^{i\theta(z)/2}=m_t(z)\,e^{i\theta_t(z)},
\ee
where $\beta_T$ is the angle between $h_1$ and $h_2$ at $T_c$,
i.e.~$\tan\beta_T=v_2/v_1$, which is less, but rather close to 1.
The top Yukawa coupling
$y_t$ is chosen such that the top mass at zero temperature is 173 GeV.
In general, the change in $\theta_2(=\theta_t)$ along the bubble wall is given
by $\Delta \theta_2= \Delta \theta/(1+\tan^2\beta_T)$, assuming that
$\tan\beta_T$ is constant along the wall \cite{HJLS99}. So there is an additional
suppression for  $\Delta \theta_2$ for large $\tan\beta$.
${\cal M}_t$ enters the computation of the baryon asymmetry; in particular, the
derivative of $\theta_t(z)$ induces the CP-violating the source term. 
Therefore a large value
of $\Delta\theta$ enhances the baryon asymmetry. As shown in
fig.~\ref{fig_theta_h2}b, $\Delta\theta$ strongly depends on $m_h$. 
Raising $m_h$ or $m_H$ does increase the change in $\theta$. We also find that
$\Delta\theta$ depends almost linearly on the coupling $\phi$, whereas the
influence of $\mu_3^2$ is small. Notice that  $\Delta \theta$ can be larger than
the input phase  $\phi$.

\section{Electric dipole moments}
CP violation induces electric dipole moments (EDMs).
The latest experimental limits for the neutron \cite{neutron_edm}
and electron \cite{electron_edm} EDMs at 90\% confidence level are
\bear
|d_{n}|&\leq&3.0\times 10^{-26}\;e\,{\rm cm},\\
|d_{e}|&\leq&1.6\times 10^{-27}\;e\,{\rm cm}.
\eear
In the standard model the only source of CP violation originates from the
Kobayashi-Maskawa matrix in the quark sector. Contributions to the EDMs arise
first at the three-loop level, which results in a natural suppression, several
orders of magnitude below current bounds. EDMs are therefore an ideal 
probe of new physics.

In the 2HDM, EDMs are induced by scalar--pseudoscalar mixing in the neutral
Higgs sector. The contributions to the EDMs can be computed in terms
of parameters ${\rm Im}(Z)$, which measure the degree of 
CP non-conservation and which are the imaginary parts of Higgs
fields normalization constants \cite{weinberg89}.
The four CP-violating parameters
\be
{\rm Im}(Z_{0i}),\hspace{3mm}{\rm Im}(\widetilde{Z}_{0i}),\hspace{3mm}
{\rm Im}(Z_{1i}),\hspace{3mm}{\rm Im}(Z_{2i}),
\ee
where $i$ indicates each of the four neutral Higgs bosons, 
enter the calculation of the EDMs. They can be expressed in terms of components
of the neutral Higgs mass matrix eigenvectors.
The Goldstone boson does not contribute to these factors, since the
corresponding $Z$'s are real. Thus, the sum can be restricted to
the three massive neutral bosons. Note that the parameters respect in
addition the sum rules \cite{weinberg89}
\be
\sum_i{\rm Im}(Z_{0i})=\sum_i{\rm Im}(\widetilde{Z}_{0i})
=\sum_i{\rm Im}(Z_{1i})=\sum_i{\rm Im}(Z_{2i})=0,
\ee
which means that CP violation vanishes if the masses of the neutral Higgs
bosons are degenerate.

In the 2HDM the dominant contributions to the electron EDM are two-loop
amplitudes, which were first computed by Barr and Zee \cite{BZ90}.
They demonstrated that the effect is enhanced with respect to the standard
one-loop contributions \cite{eedm1}. Further two-loop diagrams, including the
W-boson, were taken into account in investigations by Gunion and Vega
\cite{GV90}, Chang et al.~\cite{CKY91}, as well as Leigh et
al.~\cite{LPX91}. In this work we use the results of Chang et al., ignoring
some minor corrections discussed in ref.~\cite{LPX91}.
We end up with the following contributions
\bear
d_{e}/e&=&(d_{ e}/e)^{H\gamma\gamma}_{t {\rm -loop}}
+(d_{ e}/e)^{HZ\gamma}_{t {\rm -loop}}
+(d_{ e}/e)^{H\gamma\gamma}_{W {\rm -loop}}\nn\\
&&+\,(d_{ e}/e)^{HZ\gamma}_{W {\rm -loop}}
+(d_{ e}/e)^{H\gamma\gamma}_{G {\rm -loop}}
+(d_{ e}/e)^{HZ\gamma}_{G {\rm -loop}}.
\eear

When computing the EDM of the neutron one has to deal with
hadronic effects, which make its relation to the partonic
EDMs difficult. Various proposals have been made in the
literature how to perform this calculation (see ref.~\cite{PR05}
for a recent review). The dominant contributions to the neutron EDM
come from the color EDMs (CEDMs) of the constituent quarks $\tilde d_k$, $k=u,d$
\cite{GW90},
\be
{\cal L}\supset-\frac{i}{2}\tilde d_k g_s\bar\psi_k\sigma_{\mu\nu}G^{\mu\nu}\gamma_5\psi_k
=\frac{1}{2}\tilde d_k g_s\bar\psi_k\sigma_{\mu\nu}\tilde G^{\mu\nu}\psi_k,
\ee
and from Weinberg's three-gluon operator \cite{weinberg89}
\be
{\cal L}\supset\frac{1}{3}wf^{abc}G^a_{\mu\nu}\tilde G^{\nu\beta,b}G_{\beta}^{\mu,c}.
\ee
The QCD-corrected coefficients $\tilde d_{u}$, $\tilde d_d$ and $w$ are given by
2-loop calculations \cite{GW90,HKMT94,D90,C92}.
Using the results of \cite{PR05}, based on QCD sum rule techniques,
the neutron EDM reads
\be
(d_{ n}/e)(\tilde d_{ u},\tilde d_{ d})=(1\pm0.5)(0.55\tilde d_{ u}+1.1\tilde d_{ d})
\ee
and
\be
|(d_{ n}/e)(w)|=22{\rm ~MeV~}|w|.
\ee
Thus there is an error of about 50\% in $(d_{n}/e)(\tilde d_{u},\tilde
d_{d})$ and furthermore
an error of about 100\% in $(d_{n}/e)(w)$. Moreover, it is not possible to
determine the sign of $(d_{n}/e)(w)$. Fortunately, this latter contribution
turns out to be small (typically a 1\% correction). The same is true for the
contributions of the quark EDMs $d_{u}$ and $d_{d}$ (typically a 10\% correction).

Let us now 
\begin{figure}[t]
\begin{center} 
   \epsfig{file=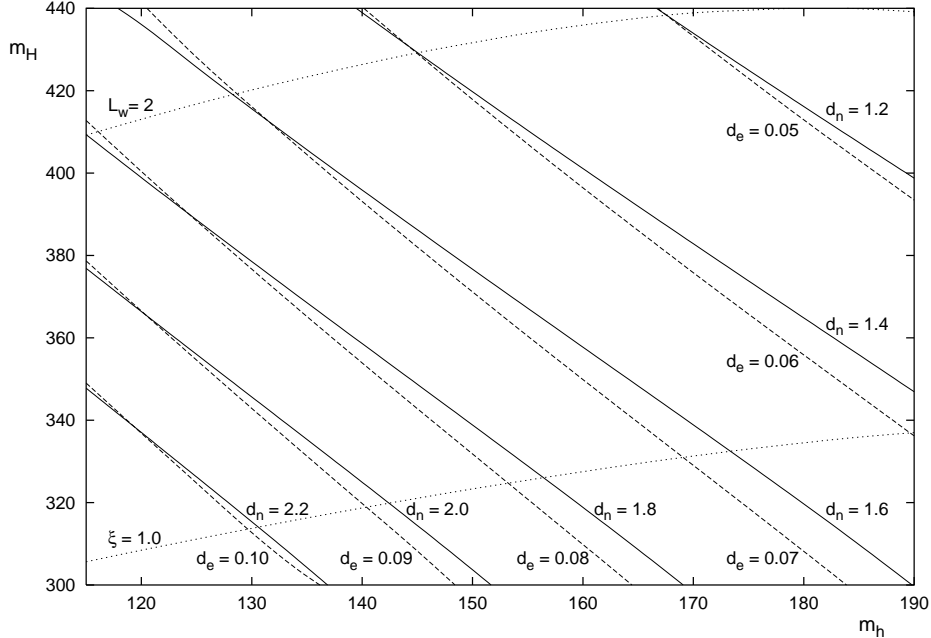,width=85mm,angle=270}
\end{center}
\caption{Lines of constant neutron (solid) and electron EDMs (dashed) for the
  set $\mu_3^2=10000\;\GeV^2$, $\phi=0.2$, $d_{n}$ is given in units of
  $10^{-26}\;e\,{\rm cm}$, $d_{e}$ in units of $10^{-27}\;e\,{\rm cm}$, and
  Higgs masses in GeV. The lower dotted line indicates the bound $\xi=1$, the
  upper one $L_{\rm w}=2$.}
\label{fig_EDMs}
\end{figure}
discuss the relevance of the electron and neutron EDMs for the 2HDM.
One finds that in the analyzed parameter region the value of $d_{e}$ is about
five to thirty times smaller than the experimental limit of $1.6\times 10^{-27}\;
e\,{\rm cm}$. Thus, there emerges no additional constraint on the parameters. 
Let us focus on the importance of the different contributions to $d_{e}$ and
on the dependence of $d_{e}$ on the input parameters. Since an
EDM arises because of CP violation, we expect a larger value for an increasing
CP phase $\phi$. Indeed we find that $d_{e}$ approximately doubles if we
change $\phi$ from 0.2 to 0.4. Also raising $\mu_3^2$
enhances $d_{e}$. Concerning the single contributions to $d_{e}$ the 
largest ones originate from the top- and $W$-loops, with 
$(d_{e})_{W {\rm -loop}}\!>\!0$ whereas
$(d_{e})_{t {\rm -loop}}\!<\!0$. The
absolute value of $(d_{e})_{t {\rm -loop}}$ is somewhat smaller, but of similar
magnitude as $(d_{e})_{W {\rm -loop}}$. So the sum is a
factor of about 5--10 smaller than each individual contribution, and is then of the
same order of
magnitude as the Goldstone-loop contribution. Thus, all three parts are important
for the electron EDM. We observe this behavior in the whole analyzed parameter
region. We also investigate the dependence of $d_{e}$ on the Higgs masses.
For increasing both, $m_h$ and $m_H$, the value of $d_{e}$ decreases. 
This tendency becomes apparent in fig.~\ref{fig_EDMs} where we compare
lines of constant electron and neutron EDMs in the $m_h$--$m_H$ plane. We find
that the larger $\mu_3^2$,  the weaker is the dependence on the heavy Higgs
mass.

Similar to the electron EDM, the one of the neutron also lies 
below its experimental bound of $3.0\times 10^{-26}\;e\,{\rm cm}$
in the analyzed parameter region. But in contrast to $d_{e}$ it almost reaches
this experimental limit. However, note that $d_{n}$ has
quite a large error of about 50\%, as pointed out. 
The limit of $3.0\times 10^{-26}\;e\,{\rm cm}$ is just saturated in the case of
$\mu_3^2=20000\;\GeV^2$ and $\phi=0.4$, for small Higgs masses. However, 
since the error band is large, there actually arises
no constraint. For larger values of $\mu_3^2$ or $\phi$, the neutron EDM of
course increases and may exceed the measured bound in a wider mass range. 
The dependence of the neutron EDM on the input parameters is quite similar to
that of the electron EDM. The lines of constant $d_{n}$ run approximately
parallel to those of $d_{e}$ in the $m_h$--$m_H$ plane; the slope is just a
little flatter. We also find roughly a doubling for a change in $\phi$
from 0.2 to 0.4. The dominant contribution arises from the color
EDM of the down-quark, which is about a factor 3.5 larger than the one due to the
up-quark CEDM. The part $|d_{n}(w)|$ arising from the three-gluon operator
is roughly an order 1\% correction and can therefore be neglected. In summary, 
for the considered parameter ranges,
both the electron and neutron electric dipole moments lie below the experimental limits.
The value of $d_{e}$ is about one order of
magnitude below the observational bound, and because of the large error in the
theoretical determination of $d_{ n}$, it also does not definitely exceed the
bound set by experiments.

\section{Transport equations}
In this section we discuss the evolution of the particle distributions during
the phase transition. The CP-violating interactions of particles in the plasma 
with the bubble wall create an excess of left-handed quarks over the corresponding
antiquarks. This excess diffuses into the
symmetric phase, where the left-handed quark density biases the
sphaleron transitions to generate a net baryon asymmetry.

Using the semiclassical WKB formalism \cite{CJK00,FH06,JPT}, 
we obtain different dispersion relations
for particles and antiparticles in the space-time dependent background of
the Higgs expectation values. The dispersion relations then lead to force terms
in the transport equations. The WKB method is justified when the de Broglie
wavelength of the particles in the plasma is much shorter than the bubble wall
thickness \cite{JPT}. Hence the condition $L_{\rm w}T\gg1$ has to be satisfied 
to legitimate an
expansion in derivatives of the background Higgs fields. As
demonstrated in section \ref{PT} we find that a
large part of the parameter space does fulfill this condition.

In the 2HDM, baryogenesis is driven by top transport. So we can focus the
discussion on the case of a single Dirac fermion, with a space-time
dependent mass ${\rm Re}{\cal M}(z)+i\gamma^5 {\rm Im}{\cal M}(z)$, 
where ${\cal M}(z)=m(z)e^{i\theta(z)}$. The dispersion relation to first order
in gradients is given by \cite{FH06,PSW1}
\begin{equation} \label{disp}
E=E_0\pm\Delta E=E_0\mp s\frac{\theta'm^2}{2E_0E_{0z}},
\end{equation}
where $E_0=\sqrt{{\bf p}^2+m^2}$ and $E_{0z}=\sqrt{p_z^2+m^2}$ in terms of the
kinetic momentum. The prime denotes the derivative with respect to $z$, and the upper and
the lower sign
corresponds to particles and antiparticles, respectively. The spin factor $s=1~
(-1)$ for $z$-spin up (down) is
related to the helicity $\lambda$ by $s=\lambda\,\mbox{sign}(p_z)$. Note that
eq.~(\ref{disp}) is the dispersion relation in
a general Lorentz frame, in contrast to the one derived in ref.~\cite{CJK00}. For the group
velocity of the WKB wave-packet one obtains
\be \label{vk}
v_{gz}=\frac{p_z}{E_0}\left(1\pm s\frac{\theta'}{2}
\frac{m^2}{E_0^2E_{0z}}\right).
\ee
The semiclassical force acting on the particles,
\be \label{Fk}
F_z=-\frac{(m^2)'}{2E_0}\pm s\frac{(m^2\theta')'}{2E_0E_{0z}}
\mp s\frac{\theta'm^2(m^2)'}{4E_0^3E_{0z}},
\ee
results from the canonical equations of motion. It was the main result of ref.~\cite{FH06} 
that the expressions for the dispersion relation (\ref{disp}), the group velocity (\ref{vk}), 
and the semiclassical force (\ref{Fk}) agree with the full
Schwinger--Keldysh result \cite{PSW1}.

In the semiclassical approximation the evolution of the particle distributions $f_i$
is described by a set of classical Boltzmann
equations. We assume a planar wall moving with constant velocity $v_{\rm w}$. Hence, 
in the rest frame of the wall, the
distributions $f_i$ only depend on $z$, $p_z$ and $p=|{\bf p}|$, due to the
translational invariance parallel to the wall. For each fluid of particle type
$i$ we have
\begin{equation}\label{Boltzmann eq}
(v_{gz}\partial_z+F_z\partial_{p_z})f_i={\bf C}_i[f],
\end{equation}
without any explicit time dependence, as we are looking for a stationary solution. 
The ${\bf C}_i$ are the collision terms describing the change of the phase-space
density by particle interactions that drive the system back to equilibrium. We
introduce perturbations around the chemical and kinetic equilibrium with the
fluid-type truncation in the rest frame of the wall \cite{CJK00}
\be
f_{i}(z,p_z,p)=\frac{1}{e^{\beta[\gamma_w(E_i+v_{\rm w}p_z)-\mu_i]}\pm1}+
\delta f_{i}(z,p_z,p)
\ee
where $\beta=1/T$ and $\gamma_{\rm w}=1/\sqrt{1-v_{\rm w}^2}$, and plus (minus) 
refers to fermions (bosons). Here the chemical potentials $\mu_i(z)$ model a local 
departure from the equilibrium 
particle density and the perturbations $\delta f_{i}$ describe the movement of
the particles in response to the force. The latter do not contribute
to the particle density, i.e.~$\int d^3p~\delta f_{i}=0$. To first order in
derivatives the perturbations are CP-even and equal for particles and antiparticles. 
But to second order they have CP-even and CP-odd parts, which we treat separately, i.e.
\be
\mu_{i}=\mu_{i,1e}+\mu_{i,2o}+\mu_{i,2e},~~~~~~~~
\delta f_{i}=\delta f_{i,1e}+\delta f_{i,2o}+\delta f_{i,2e},
\ee
so that the perturbations to second order for particles differ from those for
antiparticles.

In order to compute the asymmetry in the left-handed quark density, we expand
the Boltzmann equation in gradients. In the model under consideration, the
most important particle species are top and bottom quarks, as well as the Higgs bosons. 
The other
quark flavors and the leptons can be neglected thanks to their small Yukawa
couplings. In a first step we assume baryon number conservation. We
take into account $W$-scatterings, the top Yukawa interaction, the strong
sphalerons, the top helicity flips and Higgs number violation with rates
$\Gamma_W$, $\Gamma_y$, $\Gamma_{ss}$, $\Gamma_m$ and $\Gamma_h$,
respectively, where the latter two are only
present in the broken phase. After the left-handed quark asymmetry is computed,
the weak sphalerons, with the rate $\Gamma_{ws}$, convert it into a baryon
asymmetry. 

We follow the computation and notation presented in ref.~\cite{FH06}. 
We weight the Boltzmann equations with 1 and $p_z/E_0$, and
perform the momentum average. Accordingly ``plasma velocities" appear in the following, 
which are defined as $u_i\equiv \left<(p_z/E_0)\delta f_i\right>$. 
We end up with the transport equations for
chemical potentials of left-handed SU(2) doublet tops $\mu_{t,2}$, left-handed 
SU(2) doublet bottoms $\mu_{b,2}$, left-handed SU(2) singlet tops $\mu_{t^c,2}$, 
Higgs bosons $\mu_{h,2}$, and the corresponding
plasma velocities
\bear \label{mus}
3v_{\rm w}K_{1,t}\mu_{t,2}'+3v_{\rm w}K_{2,t}(m_t^2)'\mu_{t,2}+3u_{t,2}'&&
\nonumber\\
-3\Gamma_y(\mu_{t,2}+\mu_{t^c,2}+\mu_{h,2})-6\Gamma_m(\mu_{t,2}+\mu_{t^c,2})
-3\Gamma_W(\mu_{t,2}-\mu_{b,2})&&
\nonumber\\
-3\Gamma_{ss}[(1+9K_{1,t})\mu_{t,2}+(1+9K_{1,b})\mu_{b,2}+(1-9K_{1,t})\mu_{t^c,2}]
&=&0
\nonumber\\[.5cm]
3v_{\rm w}K_{1,b}\mu_{b,2}'+3u_{b,2}'&&
\nonumber\\
-3\Gamma_y(\mu_{b,2}+\mu_{t^c,2}+\mu_{h,2})
-3\Gamma_W(\mu_{b,2}-\mu_{t,2})&&
\nonumber\\
-3\Gamma_{ss}[(1+9K_{1,t})\mu_{t,2}+(1+9K_{1,b})\mu_{b,2}+(1-9K_{1,t})\mu_{t^c,2}]
&=&0
\nonumber\\[.5cm]
3v_{\rm w}K_{1,t}\mu_{t^c,2}'+3v_{\rm w}K_{2,t}(m_t^2)'\mu_{t^c,2}+3u_{t^c,2}'&&
\nonumber\\
-3\Gamma_y(\mu_{t,2}+\mu_{b,2}+2\mu_{t^c,2}+2\mu_{h,2})-6\Gamma_m(\mu_{t,2}+\mu_{t^c,2})&&
\nonumber\\
-3\Gamma_{ss}[(1+9K_{1,t})\mu_{t,2}+(1+9K_{1,b})\mu_{b,2}+(1-9K_{1,t})\mu_{t^c,2}]
&=&0
\nonumber\\[.5cm]
4v_{\rm w}K_{1,h}\mu_{h,2}'+4u_{h,2}'&&
\nonumber\\
-3\Gamma_y(\mu_{t,2}+\mu_{b,2}+2\mu_{t^c,2}+2\mu_{h,2})-4\Gamma_h\mu_{h,2}&=&0
\eear
\bear \label{us}
-3K_{4,t}\mu_{t,2}'+3v_{\rm w}\tilde K_{5,t}u_{t,2}'+3v_{\rm w}\tilde K_{6,t}(m_t^2)'u_{t,2}
+3\Gamma^{\rm tot}_tu_{t,2}&=&S_t
\nonumber\\[.5cm]
-3K_{4,b}\mu_{b,2}'+3v_{\rm w}\tilde K_{5,b}u_{b,2}'+3\Gamma^{\rm tot}_bu_{b,2}&=&0
\nonumber\\[.5cm]
-3K_{4,t}\mu_{t^c,2}'+3v_{\rm w}\tilde K_{5,t}u_{t^c,2}'
+3v_{\rm w}\tilde K_{6,t}(m_t^2)'u_{t^c,2}+3\Gamma^{\rm tot}_tu_{t^c,2}&=&S_t
\nonumber\\[.5cm]
-4K_{4,h}\mu_{h,2}'+4v_{\rm w}\tilde K_{5,h}u_{h,2}'+4\Gamma^{\rm tot}_hu_{h,2}&=&0.
\eear
Here the second-order perturbations label the difference between particles and
antiparticles, i.e.~$\mu_2=\mu_{2o}-\bar{\mu}_{2o}$ and
$u_2=u_{2o}-\bar{u}_{2o}$. On the r.h.s., $S_t$ denotes the source term
of the top quark,
\begin{equation}\label{Source_t}
S_t=-v_{\rm w}K_8(m_t^2\theta_t')'+v_{\rm w}K_9 \theta_t'm_t^2(m_t^2)'.
\end{equation}
The source term of the bottom quark, which is suppressed by $m_b^2/m_t^2\sim 10^{-3}$,
has been neglected. The Higgs bosons do not have a source term to second order
in gradients.
The various thermal averages $K_i$ in eqs.~(\ref{mus}), (\ref{us})
and (\ref{Source_t}) are defined similarly to ref.~\cite{FH06}.
We include the position dependence of the $K_i$.
The damping of $u_{i,2}$ can be approximated by the total
interaction rate, $\Gamma_i^{\rm tot}$. 
In the numerical
evaluations we have included a term $3\Gamma_W(u_{t,2}-u_{b,2})$ which
affects results only at the few percent level.
Contrary to the transport equations in
ref.~\cite{FH06}
we have doubled the degrees of freedom of the Higgs bosons to account for the
second Higgs doublet in the model.

Using baryon number conservation, the chemical potential of left-handed quarks
can be expressed in terms of the solutions of the transport equations $\mu_{t,2}$,
$\mu_{b,2}$ and $\mu_{t^c,2}$,
\begin{eqnarray}
\mu_{B_L}&=&\mu_{q_1,2}+ \mu_{q_2,2}+\frac{1}{2}(\mu_{t,2}+\mu_{b,2})\nn\\
&=&\frac{1}{2}(1+4K_{1,t})\mu_{t,2}+\frac{1}{2}(1+4K_{1,b})\mu_{b,2}
-2K_{1,t}\mu_{t^c,2}.
\end{eqnarray}
Now, in a second step, the weak sphalerons convert the left-handed quark number
into a baryon asymmetry.

\section{The baryon asymmetry}
The baryon asymmetry is obtained by \cite{CJK00}
\begin{equation} \label{eta1}
\eta_B=\frac{n_B}{s}=\frac{405\Gamma_{ws}}{4\pi^2v_{\rm w}g_*T}\int_0^{\infty}
dz ~\mu_{B_L}(z)e^{-\nu z}.
\end{equation}
$\Gamma_{ws}$ is the weak sphaleron rate, which is only present in the
symmetric phase, and $g_*=106.75$ is the effective
number of degrees of freedom in the plasma. The exponent
$\nu=45\Gamma_{ws}/(4v_{\rm w})$ 
accounts for the relaxation of the baryon number in case of a slowly moving
wall.

In our evaluation we use the values $\Gamma_{ws}=1.0\times10^{-6}$ for
the weak sphaleron rate \cite{Mws}, $\Gamma_{ss}=4.9\times10^{-4}T$ for the
strong sphaleron rate \cite{Mss}, $\Gamma_y=4.2\times10^{-3}T$ for the top
Yukawa rate \cite{HN95}, $\Gamma_m=m_t^2(z,T)/(63T)$ for the top
helicity flip rate \cite{HN95}, and $\Gamma_h=m_W^2(z,T)/(50T)$ for the Higgs
number violation rate \cite{HN95}. Furthermore the total interaction rate
can be expressed by the diffusion constant, $\Gamma_i^{\rm
  tot}=(D_iK_{1,i})/K_{4,i}$, where the quark diffusion constant is
given by $D_q=6/T$ \cite{HN95} and the Higgs diffusion constant by $D_h=20/T$
\cite{CJK00}. The finite $W$-scattering rate we approximate as
$\Gamma_W=\Gamma^{\rm tot}_h$. The bottom quark and the Higgs bosons are taken as
massless.

\begin{figure}[t]
\begin{center}
   \epsfig{file=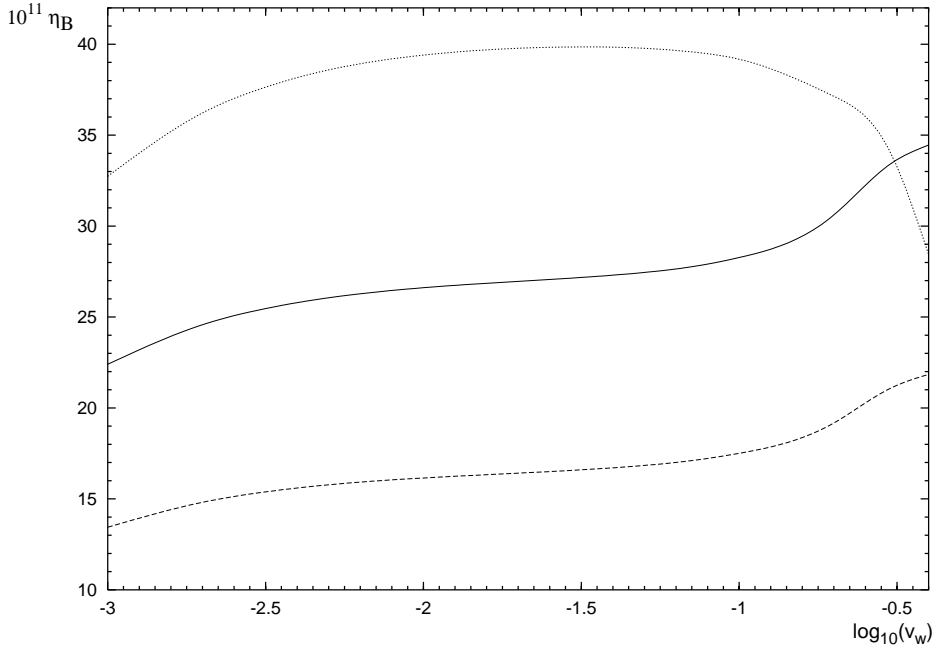,width=85mm,angle=270}
\end{center}
\caption{The solid line
  represents $\eta_B$ as a function of the wall velocity for $m_h=125$
  GeV, $m_H=350$ GeV, $\mu_3^2=10000~\GeV^2$ and $\phi=0.2$. This parameter
  setting determines $L_{\rm w}=4.5/T$ and $\xi=1.6$. The dashed line would be the
  asymmetry when we substitute $E_{0z}\rightarrow E_0$ in the dispersion
  relations. The dotted curve corresponds to the case where the $W$-scatterings
  are in equilibrium.
}
\label{2HDM-plot}
\end{figure}

Fig.~\ref{2HDM-plot} displays the baryon asymmetry as a function of the wall
velocity $v_{\rm w}$ for one typical parameter set. 
The solid line indicates the solution when using the full set of transport
equations (\ref{mus}) and (\ref{us}). If we resubstitute $E_{0z}\rightarrow E_0$
in the dispersion relation (\ref{disp}), the group velocity (\ref{vk}) and the
semiclassical force (\ref{Fk}), i.e.~going back to these quantities as determined
in ref.~\cite{CJK00}, the resulting baryon asymmetry is substantially
reduced (dashed line). This confirms the recent result that performing the boost
back to a general Lorentz frame has a sizable effect and should not be neglected
\cite{FH06}. In addition we have improved the transport equations by keeping a
finite $W$-scattering rate. If these interactions were in equilibrium, $\eta_B$
would be considerably overestimated for $v_{\rm w}\lesssim 0.1$ (dotted
curve). We could also show that taking the Higgs bosons into account or not does 
not play a significant role. The same holds for the source terms
proportional to the first-order perturbations $\mu_{t,1}$ and $u_{t,1}$, which we
have neglected in the current paper. Their effect on the total baryon asymmetry is less 
than $10\%$ in the model under consideration.

\begin{figure}
\begin{center}
   \epsfig{file=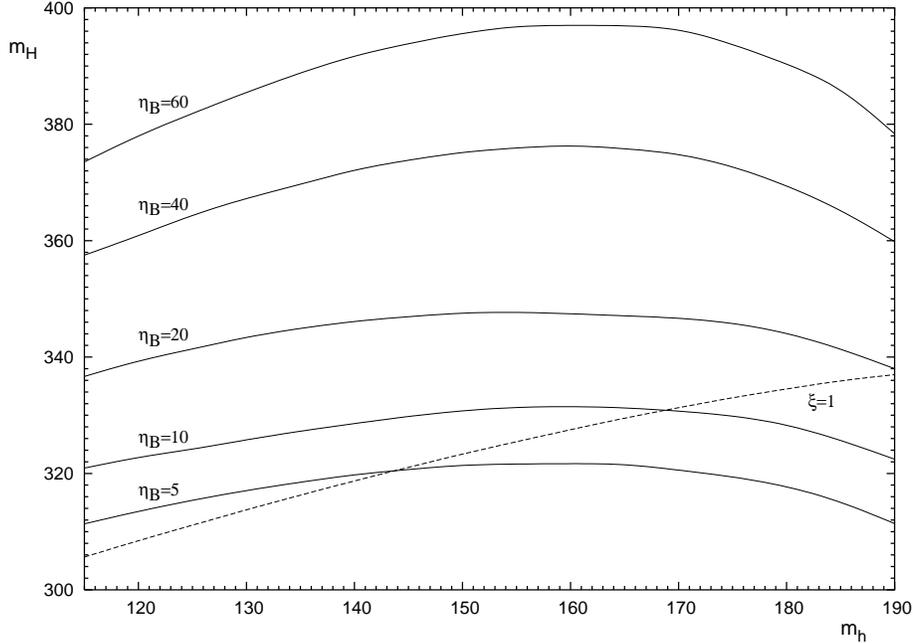,width=85mm,angle=270}
\end{center}
\caption{Contours of constant $\eta_B$ in the $m_h$--$m_H$ plane for $\mu^2_3=10000~\GeV^2$ and $\phi=0.2$. The Higgs masses
  are given in units of GeV and $\eta_B$ in units of $10^{-11}$.
}
\label{Kontur}
\end{figure}

Let us finally discuss the dependence of $\eta_B$ on the Higgs masses. 
Fig.~\ref{Kontur} shows contours of constant baryon asymmetry in the
$m_h$--$m_H$ plane, where we have fixed the parameters $\mu_3^2=10000~\GeV^2$ and
$\phi=0.2$. For each mass combination we determine all  quantities relevant for the phase
transition, such as $\xi$, $\tan\beta_T$, $L_{\rm w}$, $\theta_{\rm sym}$ and
$\theta_{\rm brk}$ to put them into the transport equations. 
There is only a mild $v_{\rm w}$ dependence of the baryon asymmetry 
(cf.~fig.~\ref{2HDM-plot}), so we
consider only one wall velocity, $v_{\rm w}=0.1$. In addition, the
($\xi$=1)-contour of fig.~\ref{fig_10000_0.2} is also shown for orientation. As we increase
$m_H$, leaving  $m_h$ fixed, the asymmetry becomes larger. 
This behavior results from the $m_t^2\sim\xi^2$ dependence of the top source term.
Accordingly the baryon asymmetry becomes larger for a stronger phase transition.
If we increase $m_h$, leaving 
the heavy Higgs mass fixed, $\eta_B$ becomes smaller and reaches a minimum at
$m_h\approx$ 150--160 GeV, similar to the behavior of $L_{\rm w}$. 
But in general there is only a minor dependence on
the light Higgs mass. 
In this parameter setting it is possible to generate the
observed baryon asymmetry for a heavy Higgs mass between 320 and 330 GeV and a
light Higgs mass up to 160 GeV. Since $\eta_B$ is more or less proportional to
the CP-violating phase $\phi$, the measured value can also be explained for
other  values of the parameters if we adjust $\phi$. Then the heavy Higgs
mass should be somewhat larger.

Comparing figs.~\ref{fig_EDMs} and \ref{Kontur}, 
we can use the baryon asymmetry to predict
the EDMs. We see that for $\xi\sim1$ and $m_{h}=115$ GeV the neutron EDM
is a factor of about 2 above the experimental bound, which including the theoretical
uncertainties is marginally tolerable. The electron EDM is a factor of about 5 below
the experimental bound. Moving along the ($\xi=1$)-contour to the largest Higgs mass
of 190 GeV, we find  $|d_{e}|\sim0.03\times 10^{-27}e~{\rm cm}$ and
$|d_{n}|\sim0.7\times 10^{-26}e~{\rm cm}$. Finally, taking $m_{h}=190$ GeV
and $m_{H}=400$ GeV, we find  $|d_{e}|\sim3\times 10^{-30}e~{\rm cm}$ and
$|d_{n}|\sim0.09\times 10^{-26}e~{\rm cm}$. So the experimental bound
on the neutron EDM starts to cut into the parameter space of the model. Improving the
bound by an order of magnitude would probe the larger part of the parameter space. 
The electron EDM is typically one to two orders of magnitude below the current bound.

In this paper we focused on the case $\tan\beta=1$. As we discussed in the context
of eq.~(\ref{M}), larger values of $\tan\beta$ will lead to a smaller value of the baryon
asymmetry, since the change in $\theta$ is then mostly due to a change in  $\theta_1$
rather than $\theta_2$.  Extrapolating from the example of fig.~\ref{Kontur}, we estimate
that for $\tan\beta\gsim10$, successful baryogenesis should in any case be in
conflict with the EDM bounds. It would be interesting to check this issue by direct
evaluations.

So there exists a wide range of realistic parameters where the computation of
$\eta_B$ is under control, and which yields the observed baryon asymmetry.

\section{Conclusions}
We have studied electroweak baryogenesis in the 2HDM, focusing on the case of
$\tan\beta=1$ and degenerate extra Higgs states. Evaluating the thermal
Higgs potential in the one-loop approximation, we find a first-order phase
transition, which is strong enough to avoid baryon number washout. This is achieved
by the loop-contributions of the extra Higgs states, provided they are sufficiently
strongly coupled. Taking $\mu_3^2=10000~\GeV^2$, this happens for a
heavy Higgs mass $m_H\gsim300$~GeV. The mass of the light, SM-like Higgs, 
$m_h$, can be up to 200 GeV, or even larger. The Higgs potential allows the
introduction of a single 
CP-violating phase, which has only a minor impact on the strength of the phase
transition. In our example, if $m_H$ reaches about 500 GeV, the phase
transition becomes very strong, while the perturbative description starts to
break down. These findings are in agreement with those of ref.~\cite{CL96}.

We have computed the properties of the phase boundary. The walls are typically
thick, but the width decreases with larger $m_H$ from $L_{\rm w}~\sim15T_c^{-1}$ 
to about $2T_c^{-1}$. We also compute the profile of the relative complex phase
between the two Higgs vevs, which changes by an amount $\Delta\theta$ between
the broken and the symmetric phase. 

This phase shift leads to a CP-violating source term for the top quark, which
drives the generation of the baryon asymmetry. We compute the source term
in the WKB approximation and solve the resulting transport equations, using the
formalism of ref.~\cite{FH06}. We find that for typical parameter values
the baryon asymmetry is in the range of the observed value. The explicit CP phase
in the Higgs potential has to be taken between $10^{-2}$ and unity.
For larger values of $m_H$ the baryon asymmetry increases, as the phase
transition becomes stronger and the wall thinner. Our result differs from those
of ref.~\cite{CKV95}. There the baryon asymmetry in the 2HDM
was computed using the method of reflection and transmission coefficients. 
In the regime of thick walls, this method is known not to give the leading 
contribution to the baryon asymmetry, which explains the different results.
 
We have also computed the EDMs of the electron and neutron. Since there is
only a single complex phase in the model, we can predict $|d_{e}|$
and $|d_{n}|$ in terms of the baryon asymmetry and the Higgs masses. We find that 
$|d_{n}|\gsim10^{-27}e~{\rm cm}$. For the smallest allowed values of $m_h$
and $m_H$,  $|d_{n}|$ can slightly exceed the experimental bound.
Improving the neutron EDM sensitivity by an order of magnitude would
test a substantial part of the parameter space of the model. The electron EDM is
typically one to two orders of magnitude below the bound. These values are
for $\tan\beta=1$. Extrapolating our results suggests that for $\tan\beta\gsim10$,
the 2HDM cannot produce the observed baryon asymmetry without being
in conflict with the EDM constraints. In any case, the 2HDM can explain the
baryon asymmetry for a considerable range of the model parameters. 
 
It would be interesting to extend our investigations to cover the full parameter space,
in particular the case $\tan\beta>1$. Since for larger values of $m_H$
higher-order corrections to the effective potential become more and more
important, these contributions should be studied in more detail, most reliably
on the lattice. Our proposal is testable at the LHC in the sense that at least one 
Higgs state should be observed. Furthermore, CP violation may be detectable
in top pair production \cite{B94,Bd06}.
Stringent tests could be performed
at a future $e^+e^-$ linear collider, where for instance deviations in the Higgs
self-coupling could be detected \cite{hhh}. 

\section*{Acknowledgements}
We thank D. B\"odeker, J. Erdmann, M.~Laine, A. Ritz and 
S. Weinstock for valuable discussions.


\end{document}